\documentclass[]{emulateapj}

\slugcomment{}

\shorttitle{Major Dry Mergers in Coma}
\shortauthors{Cordero et al.}

\begin{document}

\title{Dry Merger Rate and Post-Merger Fraction in the Coma Cluster Core}

\author{Juan P. Cordero\altaffilmark{1}, Luis E. Campusano}
\affil{Departamento de Astronom\'ia, Universidad de Chile, Casilla 36-D,
  Santiago, Chile}
\author{Roberto De Propris}
\affil{Finnish Centre for Astronomy with ESO, University of Turku, Vaisalantie 20, Piikkio, 21500, Finland}
\author{Christopher P. Haines}
\affil{Departamento de Astronom\'ia, Universidad de Chile, Casilla 36-D,
  Santiago, Chile}
\author{Tim Weinzirl}
\affil{School of Physics and Astronomy, The University of Nottingham, University Park, Nottingham NG7 2RD}
\and
\author{Shardha Jogee}
\affil{Department of Astronomy, The University of Texas at Austin, Austin, TX 78712-1205}

\altaffiltext{1}{jcordero@das.uchile.cl}
\begin{abstract}

We evaluate the dry merger activity in the Coma cluster, using a spectroscopically
complete sample of 70 red sequence (RS) galaxies, most of which ($\sim 75 \%$) are located within 0.2$R_{200}$ ($\sim 0.5$ Mpc) from the cluster center, with data from the Coma Treasury Survey obtained with the Hubble Space Telescope. The fraction of close galaxy pairs in the sample is the proxy employed for the estimation of the merger activity. We identify five pairs and one triplet, enclosing a total of 13 galaxies, based on limits on projected separation and line-of-sight velocity difference. 
Of these systems, none show signs of on-going interaction and therefore we do not find any true merger in our sample. This negative result sets a 1-sigma upper limit of 1.5\% per Gyr for the major dry merger rate, consistent with the low rates expected in present-day clusters. Detailed examination of the images of all the RS galaxies in the sample reveals only one with low surface brightness features identifiable as the remnant of a past merger or interaction, implying a post-merger fraction below two percent. 
\end{abstract}

\section{Introduction}
\begin{figure*}
\begin{center}
\includegraphics[width=0.8\textwidth]{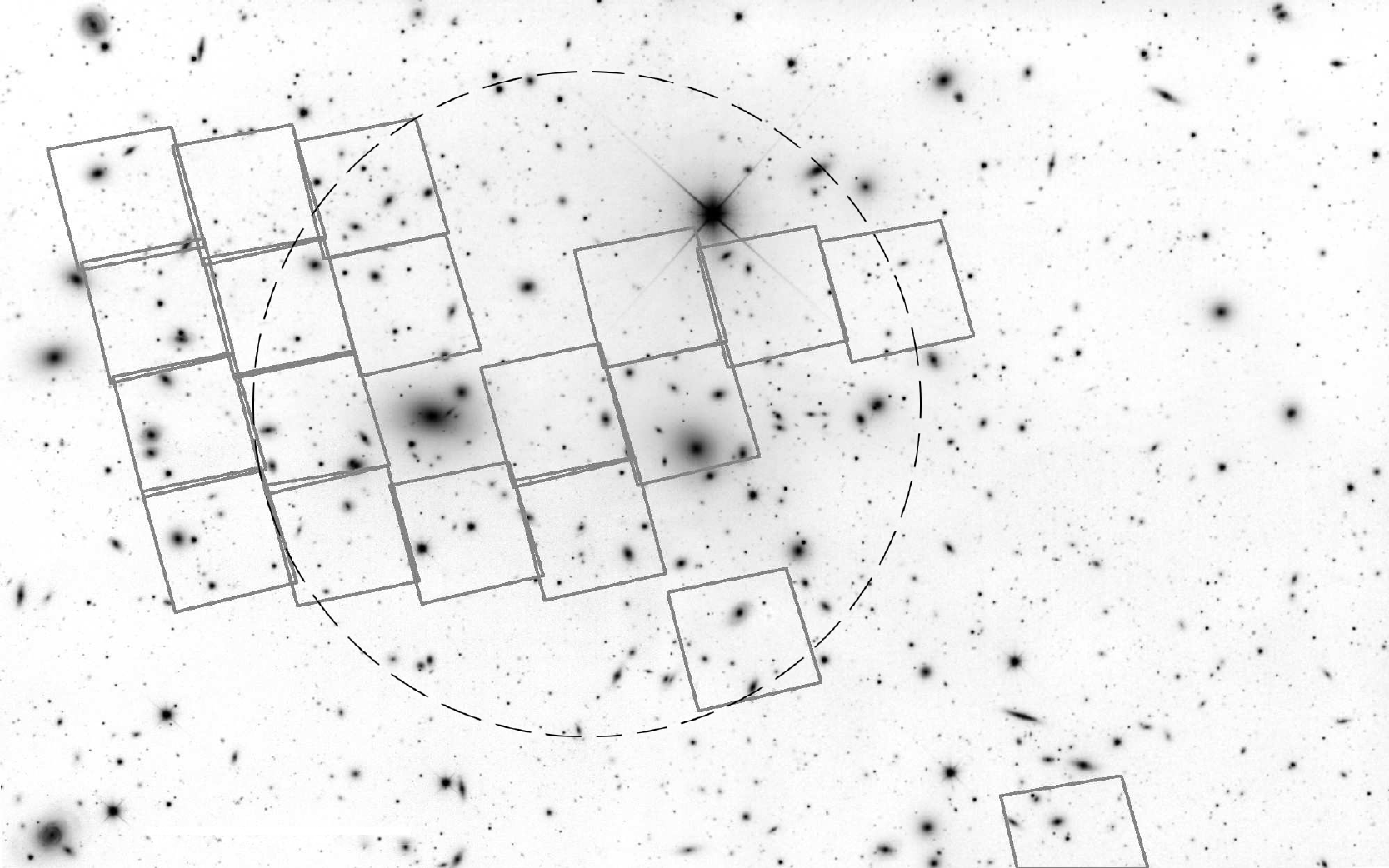}
\end{center}
\caption{The core region of Coma and the footprints of 19 HST/ACS frames, of a total of 25, distributed mainly over a $0.7 \times 0.5$ Mpc region (the circle, 0.5\,Mpc in diameter, marks the cluster center). Note that brightest galaxy, NGC4889, is not within the HST imaging.
\label{fig1}}
\end{figure*}

Mounting observational and theoretical evidence suggest
that galaxy growth proceeds through a combination
of  major mergers, \citep[e.g.,][]{2005Natur.435..629S,2009MNRAS.397..506K},
minor mergers \citep[e.g.,][]{2009ApJ...697.1971J,2011ApJ...743...87W}, cold-mode gas accretion
\citep[e.g.,][]{2009ApJ...694..396B,2009ApJ...703..785D}, and secular processes
\citep[e.g.,][]{2004ARA&A..42..603K,2005ApJ...630..837J}. Mergers are particularly important,
contributing to the stellar mass growth of galaxies, triggering star formation, inducing nuclear activity and leading to
morphological transformation. 

If major mergers are actually common, it is difficult to explain the very low scatter seen in the Fundamental Plane scaling relations even at redshifts approaching 1 and beyond \citep[e.g.,][]{2011A&A...526A..72F,2014ApJ...793L..31V}. Major dry mergers (between two gas-poor quiescent galaxies) may offer an escape from this apparent contradiction, as they are believed not to affect the scaling relations \citep[e.g., ][]{2005AAS...20710504B,2012ApJ...753...44S}.

Major mergers in present-day clusters are not expected to be frequent, as the encounter velocities between cluster galaxies are much higher than the internal velocity dispersions of the galaxies, preventing their coalescence \citep{1980ApJ...236...43A}. Therefore, the evolution in the mass function of cluster galaxies should closely follow that seen in the wider field and group populations which are continually accreted into the clusters over time \citep{2015ApJ...806..101H}. In fact, the cluster galaxy mass function appears not to have evolved significantly since $z\sim1.5$ or even earlier \citep[e.g.][]{2007AJ....133.2209D,2008ApJ...686..966M}.

Some recent results are shedding new light on the merger activity in the local universe.
Very deep imaging studies of local ($z \lesssim 0.1$) field early-type galaxies have reported that features such as broad fans, ripples, shells, streams and tidal tails are found in 50-70\% of them, pointing to recent mass assembly through dry mergers \citep{2005AJ....130.2647V,2015MNRAS.446..120D}. In a similar deep optical survey of four rich clusters at $z \lesssim 0.1$ (A119, A389, A2670, A3330), \citet{2012ApJS..202....8S} also identified such features in $\sim25$\% of RS cluster galaxies, a result particularly surprising for such environments. \citet{2012ApJS..202....8S} suggested that these faint features could be residuals of mergers which took place several Gyr ago, prior to the galaxies being accreted into the clusters themselves. \citet{2013A&A...554A.122Y} performed hydrodynamical simulations of major merging galaxies indicating that post-merger signatures could remain detectable for 3-4 Gyr.

In this Letter we consider the use of close pair fractions and image
inspection to estimate the dry merger rate of galaxies in the Coma cluster
($z=0.0231$) and the fraction of post-merger, or merger remnant, galaxies in the RS, using the
extensive available spectroscopic information and deep HST imaging
obtained with the Advanced Camera for Surveys (ACS) for the Coma Treasury
Survey (CTS). These measurements are not only useful for comparison with similar investigations of low-z clusters, but also constitute a suitable counterpart to previous studies of distant clusters ($z\sim 0.8-1.6$) where high merger fractions have been claimed based on galaxy pair counts \citep[e.g.][]{1999ApJ...520L..95V,2005ApJ...627L..25T}. The much better data quality available for local systems allows us to identify signs of on-going interactions and explore systematic effects on the determination of the merger rate.

This analysis is based on a complete spectroscopic sample of RS galaxies consisting of gas-poor elliptical and lenticular galaxies. The dry merger rate is derived from the number of close pairs that show signs of galaxy-galaxy interactions in the model subtracted images of the component galaxies. We also estimate the post-merger fraction in Coma from the number of galaxies in the complete RS sample that show remnant features from a past coalescense, for comparison with the results of \citet{2012ApJS..202....8S}.
A cosmology with $\Omega_m=0.3$, $\Omega_{\Lambda}=0.7$ and $H_0=70$ km s$^{-1}$ Mpc$^{-1}$ is adopted.

\begin{figure*}
\begin{center}
\includegraphics[width=0.65\textwidth]{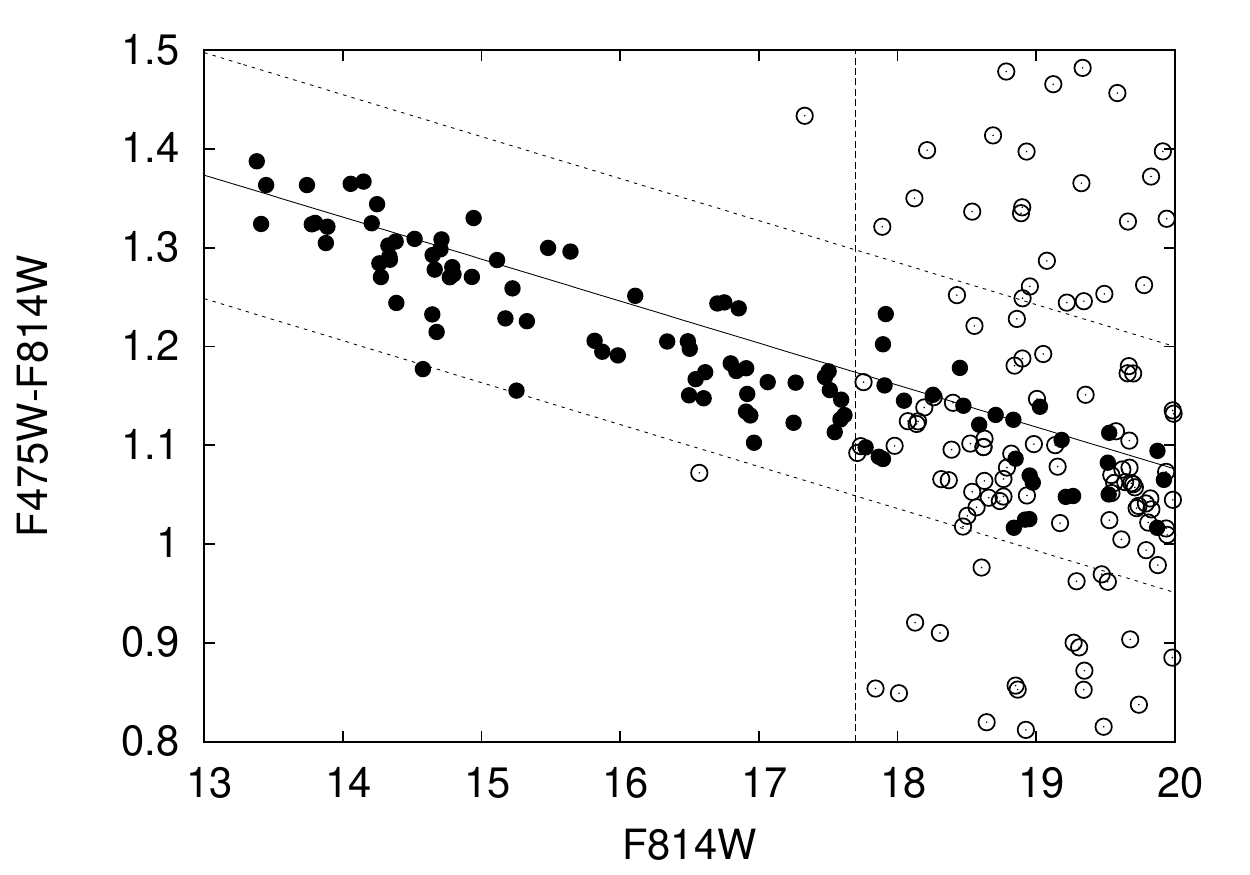}
\caption{Color-magnitude diagram for objects brighter than F814W = 20. Red sequence galaxies lie within the 5$\sigma$ region delimited by dotted lines, while the red sequence relation is indicated by the solid line and is given by F475 - F814W = -0.0425$\times$F814W + 1.916. Open circles show galaxies with unknown radial velocity, while filled black circles correspond to member galaxies with known radial velocities between 4,000 and 10,000 km s$^{-1}$. Vertical line at F814W $=$ 17.7 ($\sim 10^9 M_{\odot}$) marks the limit for a complete spectroscopic sample.}
\end{center}
\end{figure*}

\section{Data \& Sample selection}

We select Coma galaxies within the footprint (Fig. 1) of the Coma Treasury Survey
\citep[CTS -- ][]{2008ApJS..176..424C}. The angular scale for Coma is 0.472 kpc arcsec$^{-1}$. The CTS provides high quality ($0.05$\,arcsec\,pixel$^{-1}$) imaging in both F475W ($g$) and F814W ($I$) HST filters, and only the galaxies contained therein are covered by our analysis. The surface brightness limit (SBL) is estimated to be $\Sigma_{\rm F814W}\sim 26.5$\,mag\,arcsec$^{-2}$ at the 3$\sigma$ level. Nineteen pointings, out of 25, cover roughly 20 per cent of the projected area within 0.5 Mpc from the Coma center. The other 6 pointings are between 0.9 and 1.75 Mpc south-west of the cluster center. We use the available photometry from the CTS \citep{2010ApJS..191..143H} to construct the color-magnitude diagram and determine the red sequence (Fig.~2). Radial velocities are compiled from NED. One-hundred and seventy-six galaxies, ranging from F814W$\sim$13.5 to ${\sim}20$ mag and lying within 5$\sigma$ (dotted lines) of the mean relation, are considered to be RS. Out of these, 70 are brighter than F814W$=17.7$ mag, the limit to which the redshift information is 100\% complete, and have radial velocities between 4,000 and 10,000 km s$^{-1}$, the redshift limits for membership of the Coma cluster. Using the stellar mass estimates from \citet{2014MNRAS.441.3083W}, this limit equates to a threshold of  ${\sim}10^9\,{\rm M}_{\odot}$.

\begin{figure*}
\epsscale{0.5}
\flushleft
\plotone{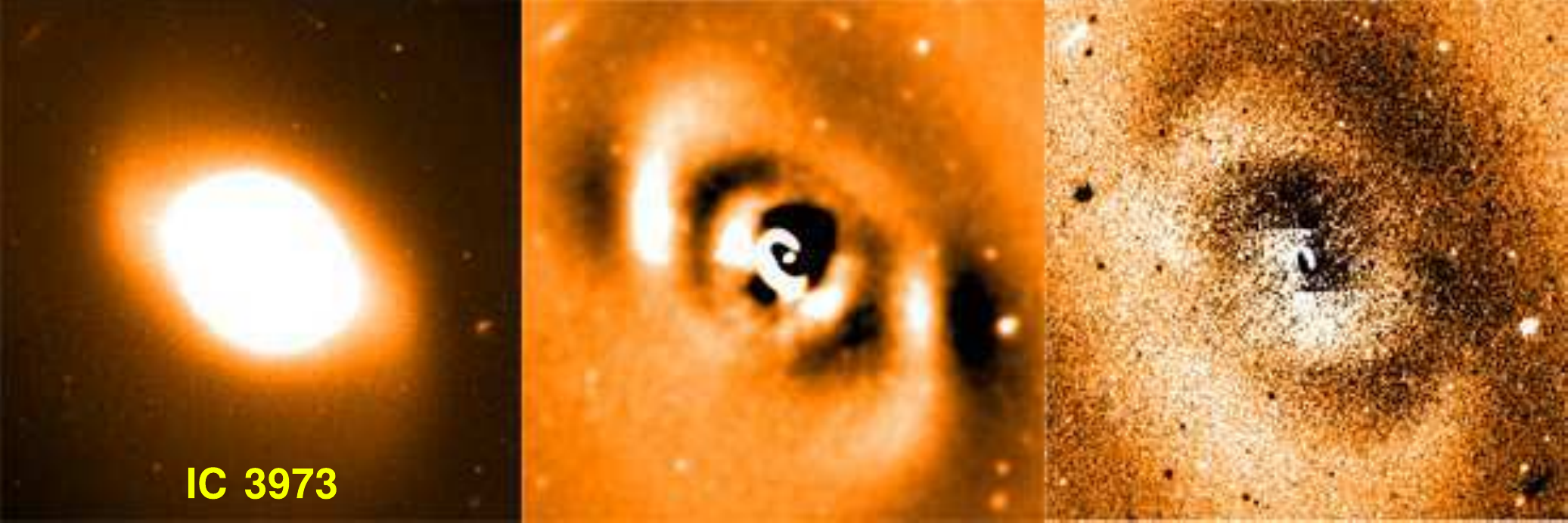}  \plotone{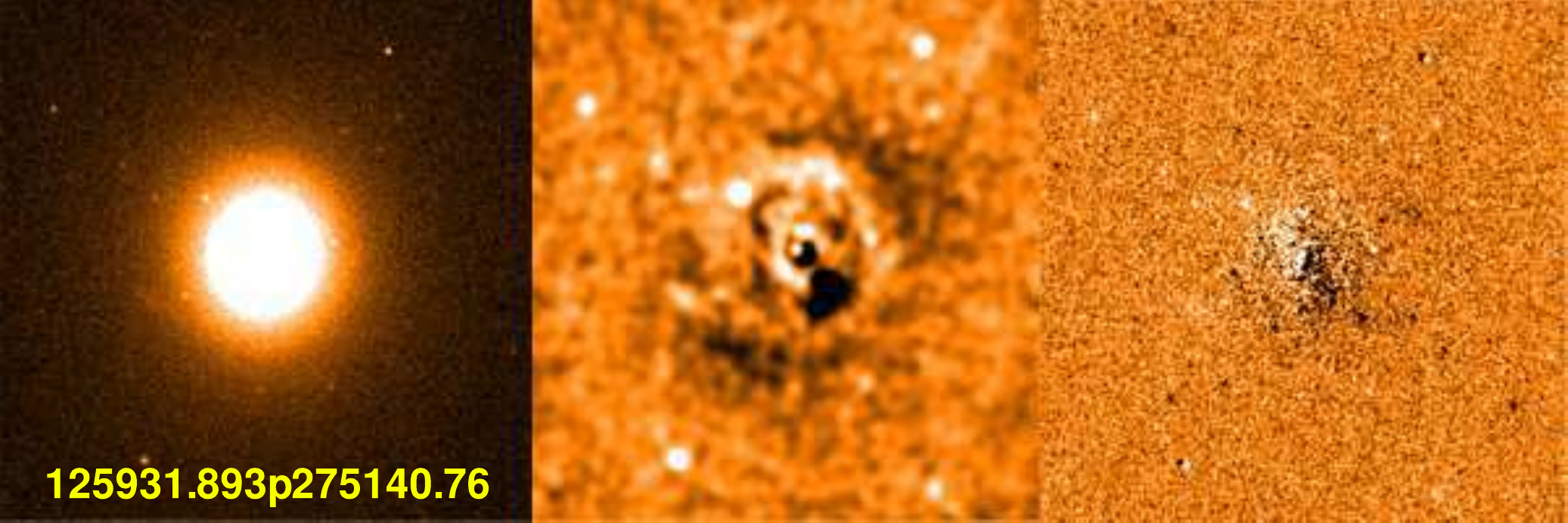} \\
\vspace{0.1cm}
\plotone{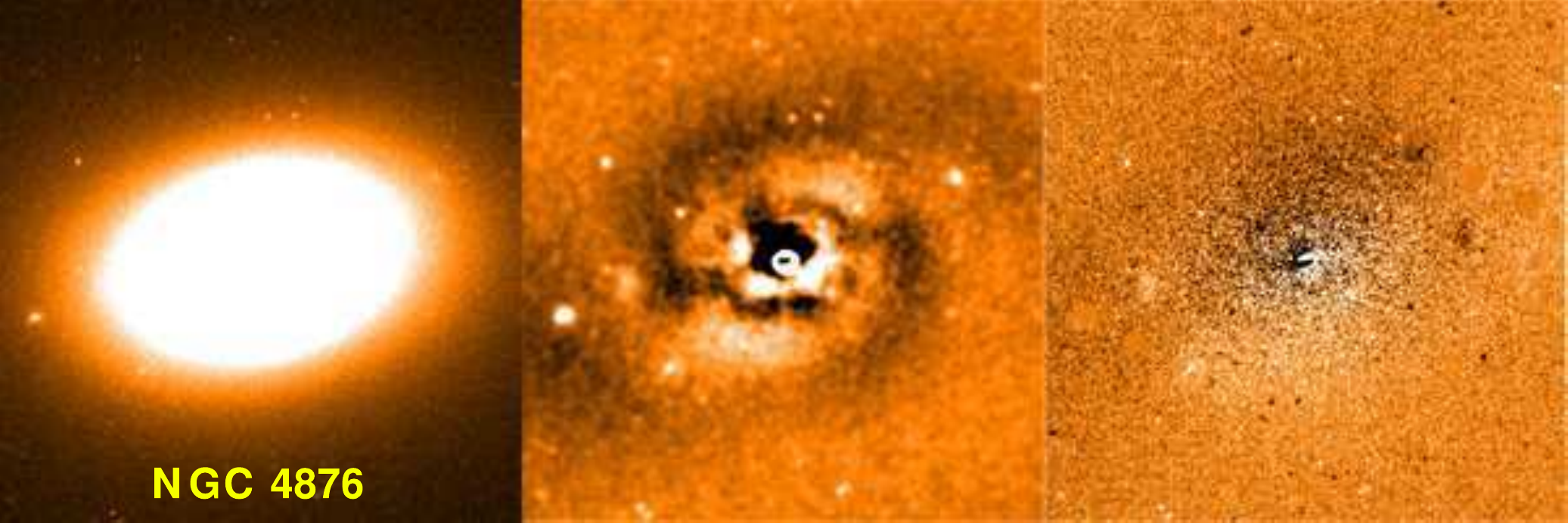}  \plotone{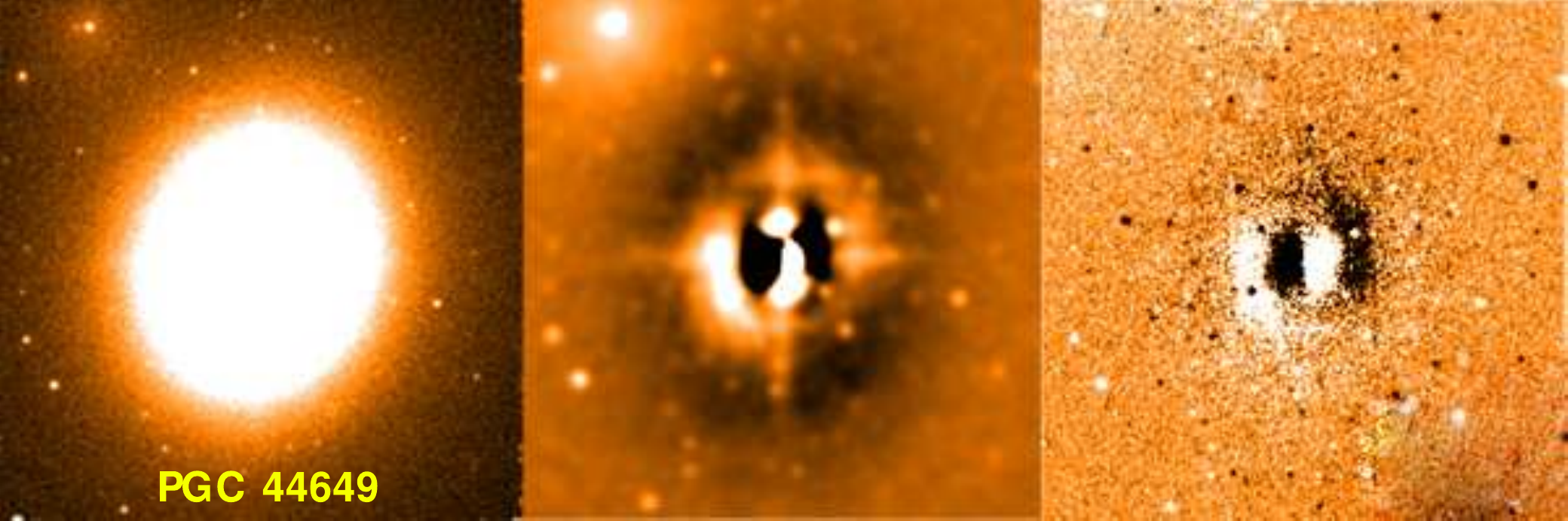} \\
\vspace{0.1cm}
\plotone{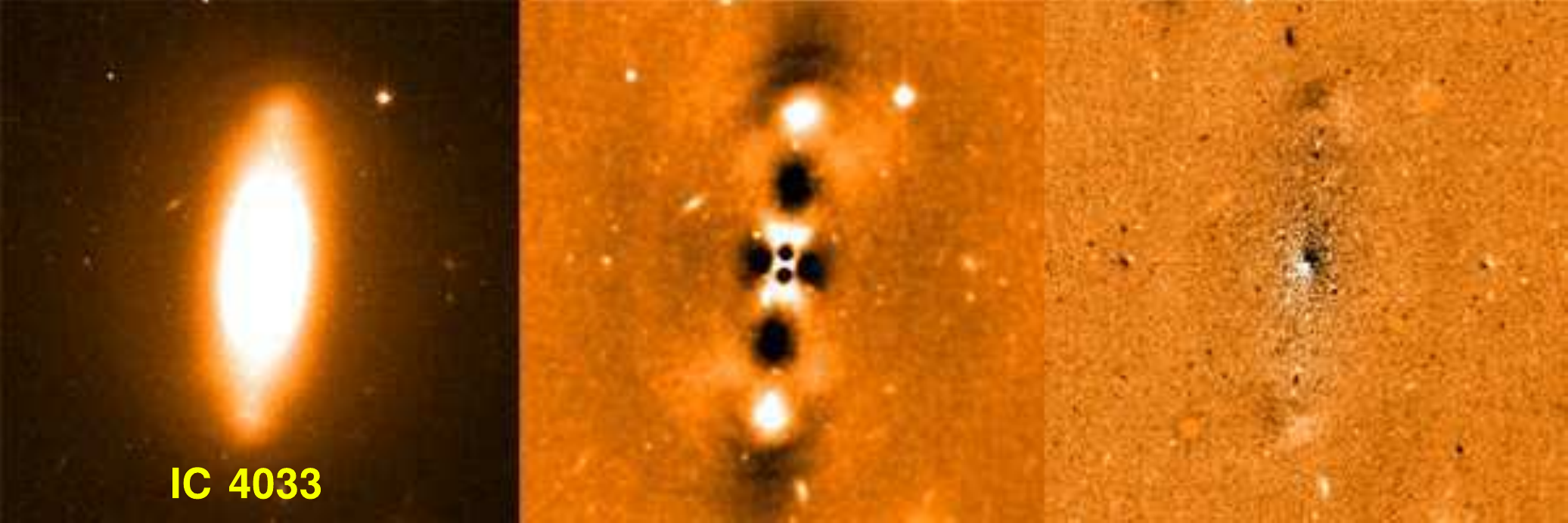}  \plotone{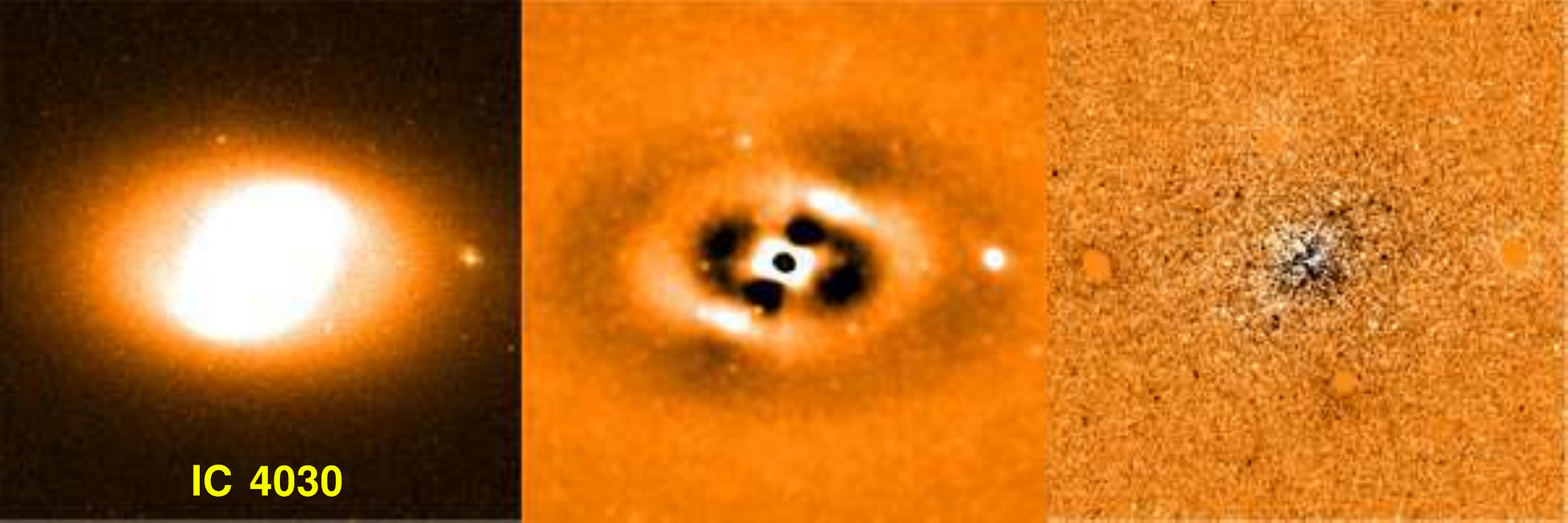} \\
\vspace{0.1cm}
\plotone{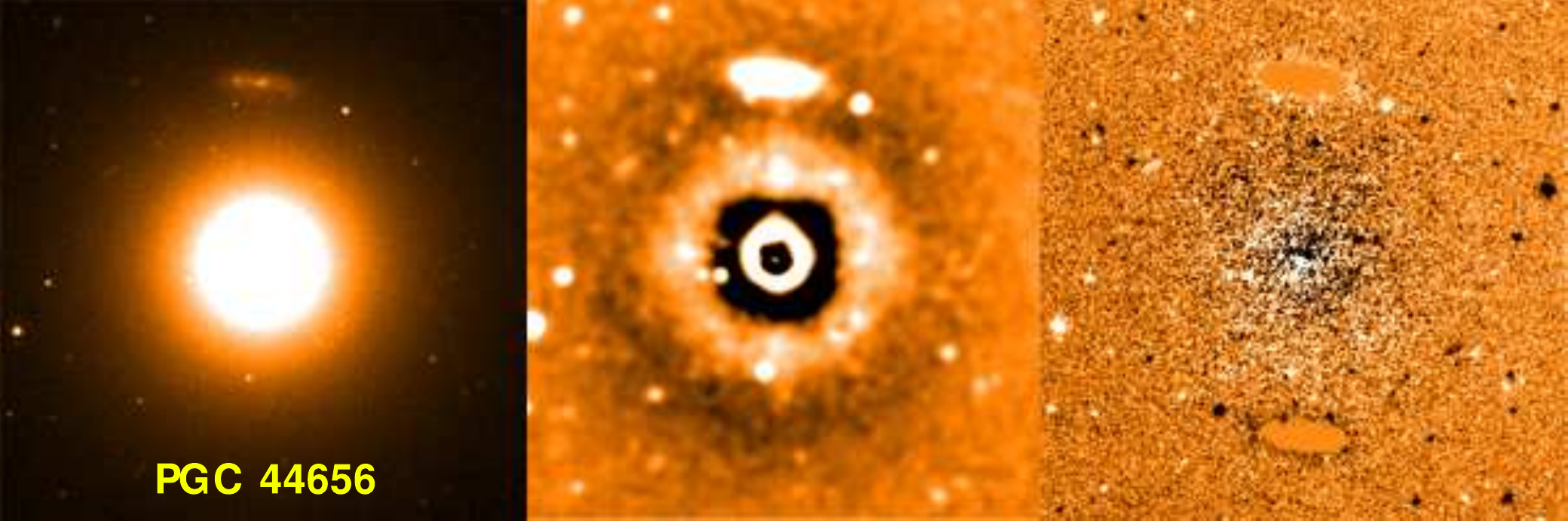}  \plotone{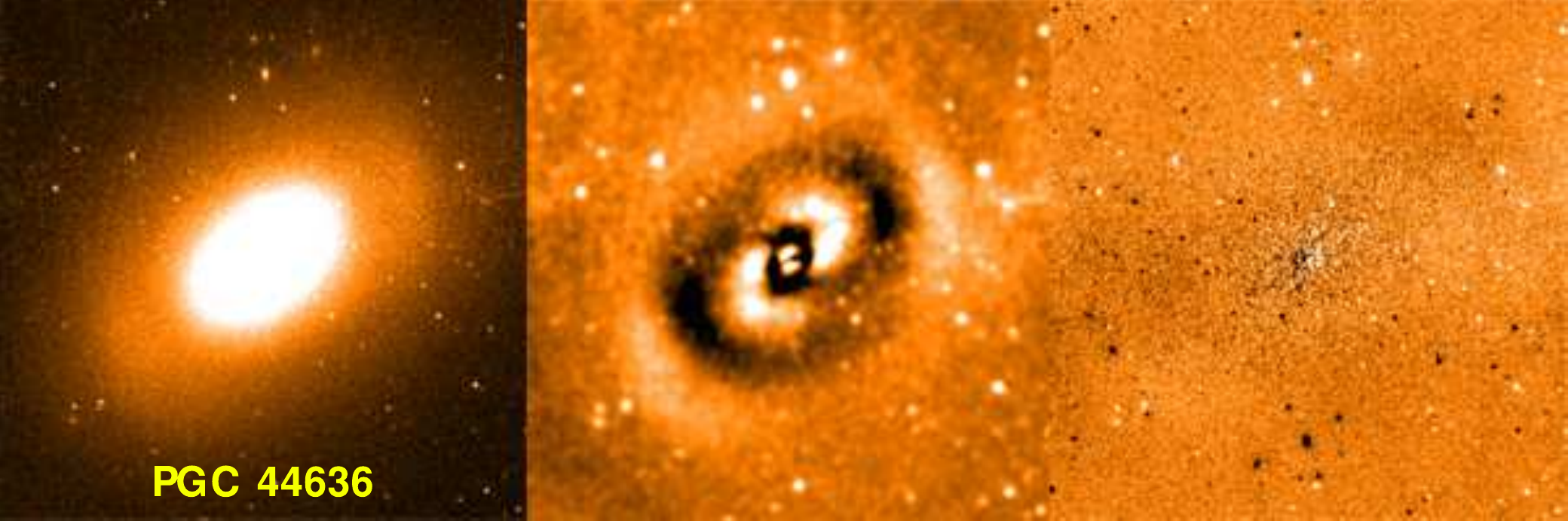} \\
\vspace{0.1cm}
\plotone{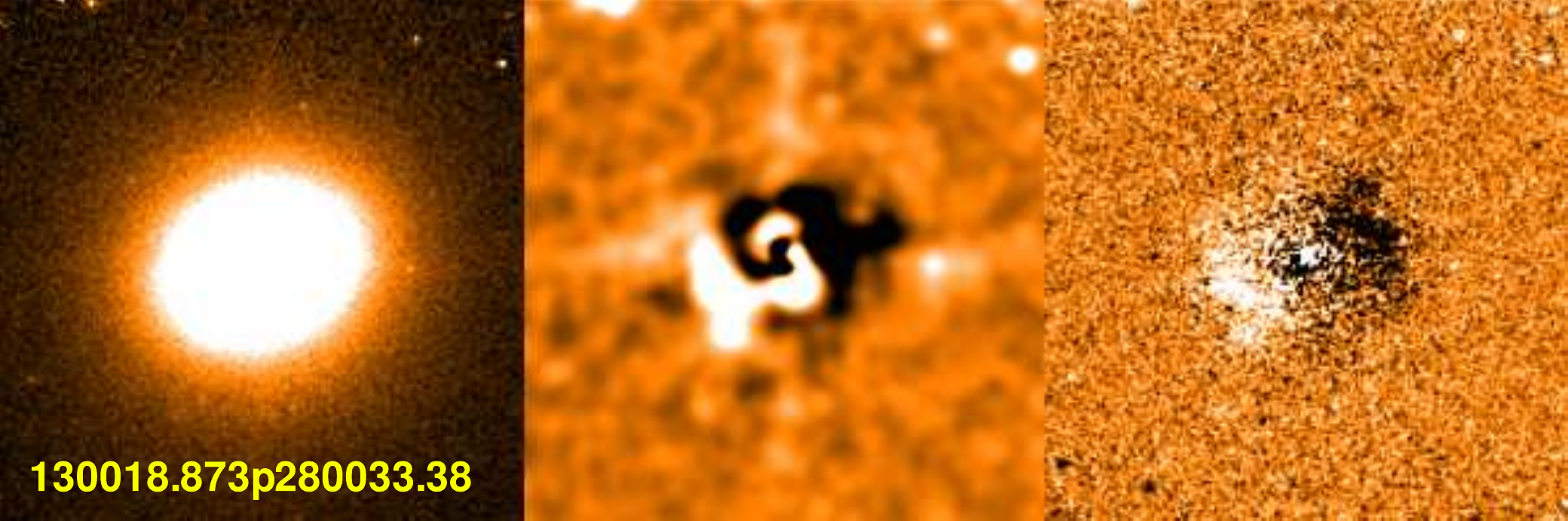}  \plotone{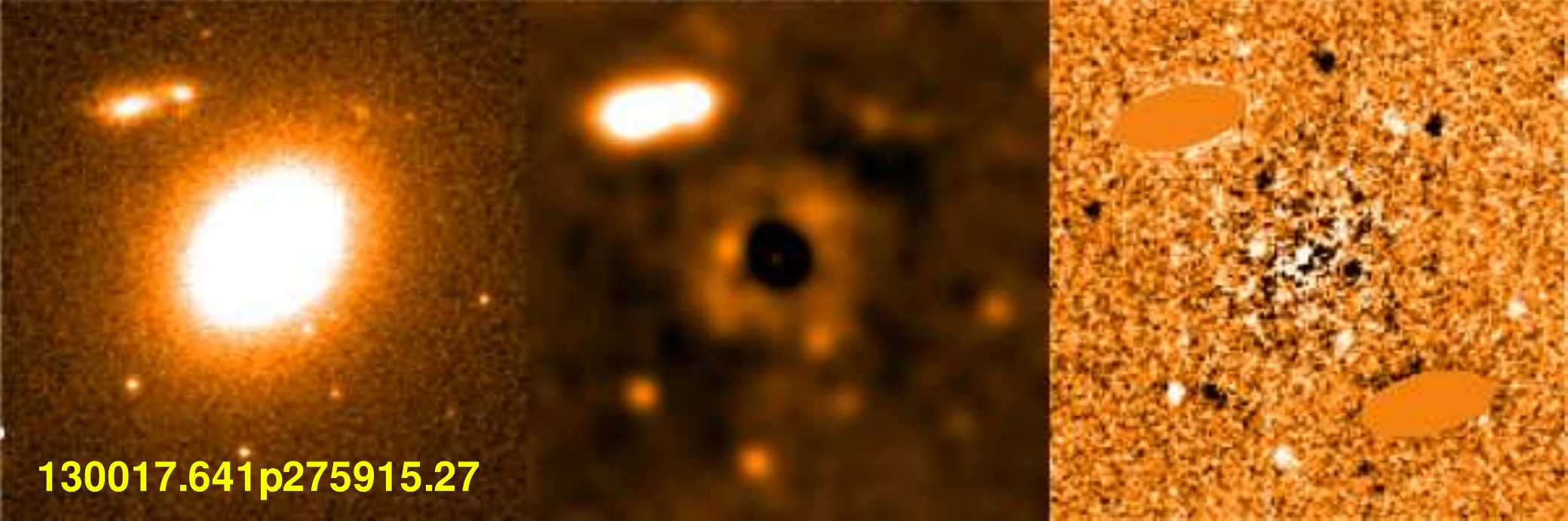} \\
\vspace{0.1cm}
\plotone{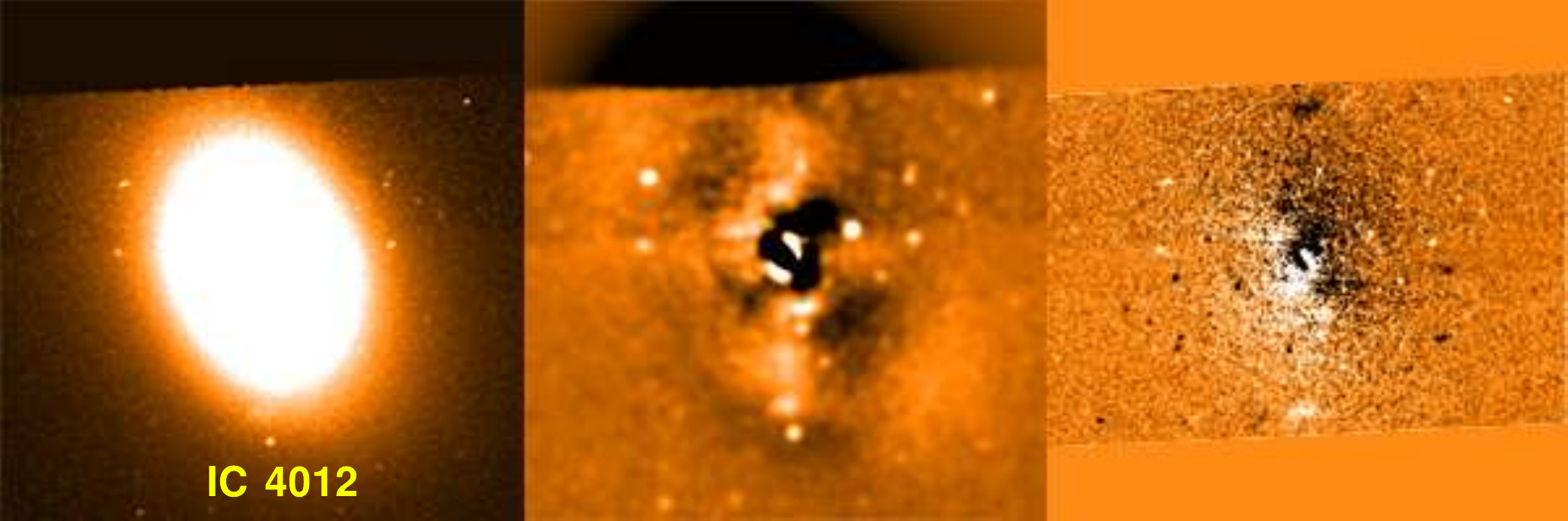}  \plotone{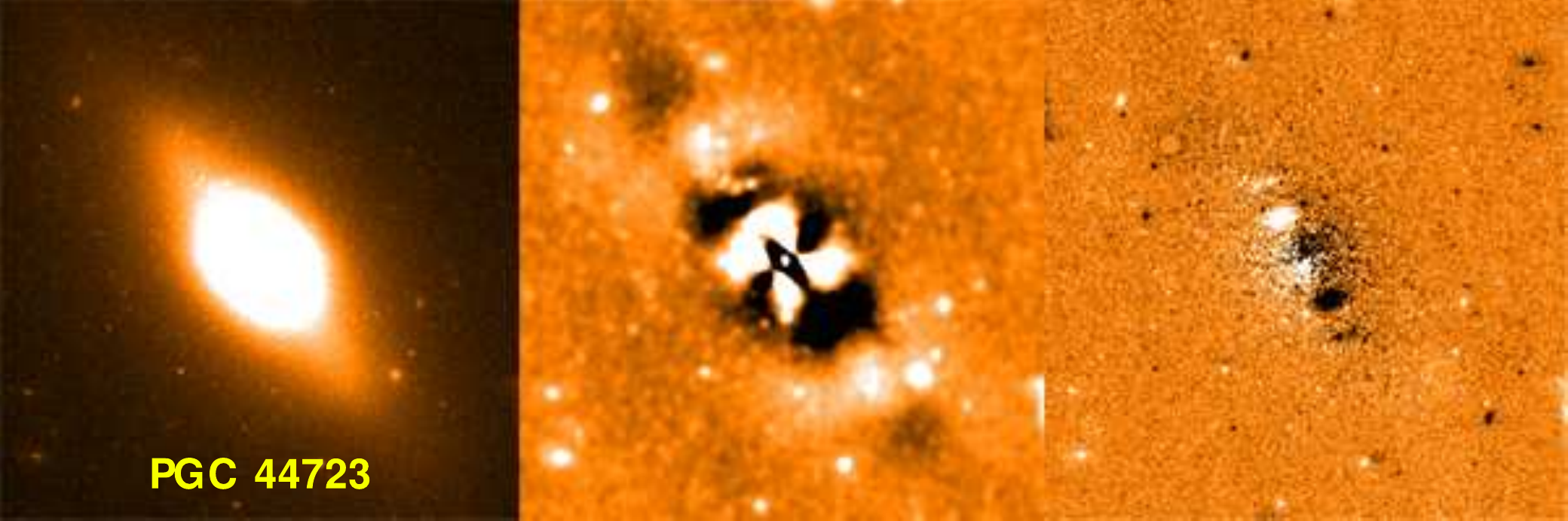} \\
\vspace{0.1cm}
\plotone{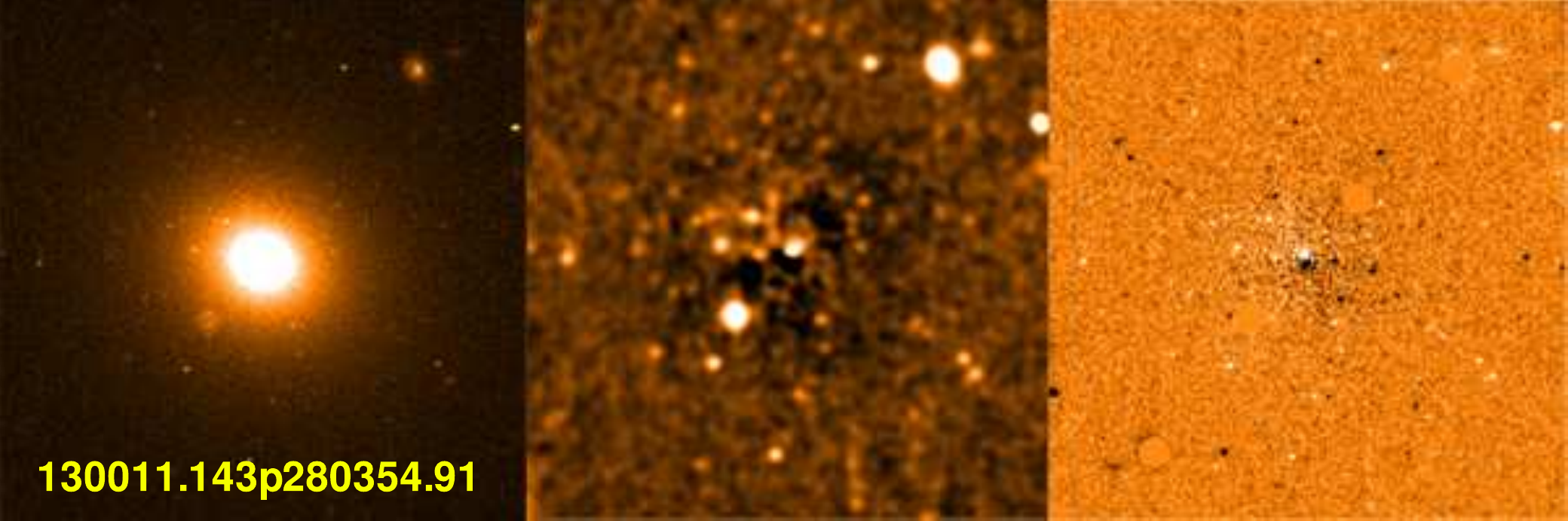}
\caption{F814W images, model and rotation residuals for the 13 galaxies in pairs/triplet are presented from left to right on each column.  The first five rows correspond to the pairs. The last three images on the bottom correspond to the galaxies in the triple system. \label{fig3}}
\end{figure*}

\section{Pairs}
\subsection{Pair selection}
\label{Pair selection}
The complete sample is searched for close pairs/triplets, which are considered as pre-mergers if they show signatures of interaction on the HST images. This replicates the approaches of \citet{1999ApJ...520L..95V}, \citet{2008ApJ...683L..17T} and \citet{2012ApJ...755...14R} in more distant clusters, but adds progressively more information to test systematic effects on the determination of merger rates in distant clusters. Additionally, asymmetry parameters are measured for these galaxies.

\begin{deluxetable*}{llccccrcc}
\tabletypesize{\scriptsize}
\tablecaption{{Galaxies belonging to selected pairs/triplets by projection and velocity proximity (see text) }\label{Table1}}
\tablewidth{0pt}
\tablehead{
 & & Stellar Mass& \colhead{F814W} & \colhead{Projected $r_s$} \vspace{-0.1cm} & \colhead{$\Delta V$} & & &\\
 \colhead{CTS ID\tablenotemark{a}} \vspace{-0.2cm}  & \colhead{Name} & & & & & \colhead{$A_{abs}$} & Sersic index & Morphology\\
 & & ($M_{\odot}$)& \colhead{(mag)} & \colhead{(kpc h$^{-1}$)} & \colhead{(km s$^{-1}$)} & & &\\ \vspace{-0.3cm}
}
\startdata
125930.824p275303.05 \vspace{-0.1cm}& IC 3973& 5.02$\times 10^{10}$ & 13.77 &  &  & 0.0419 & 3.77& S0/a\\ \vspace{-0.1cm}
 & & & & 27.70 & 135 \\
125931.893p275140.76& -& 1.96$\times 10^{9}$ & 16.91 &  &  & 0.0332 &1.88 & E\\
\\
125944.407p275444.84 \vspace{-0.1cm}&NGC 4876 &4.62$\times 10^{10}$ & 13.89 &  &  & 0.0287 & 2.96&S0 \\ \vspace{-0.1cm}
& & & & 17.25 & 222 \\
125942.301p275529.15& PGC 44649 & 1.61$\times 10^{10}$&14.93 &  &  & 0.0590 &7.49 & S0\\
\\
130028.370p275820.64  \vspace{-0.1cm} &IC 4033 &2.17$\times 10^{10}$ &14.65 &  &  & 0.0248  & 4.04& S0\\ \vspace{-0.1cm}
& & & & 19.60 & 129 \\
130027.966p275721.56 &IC 4030 & 2.11$\times 10^{10}$&14.70 &  &  & 0.0270 &4.68 & S0\\
\\
125943.721p275940.82  \vspace{-0.1cm} & PGC 44656 &2.13$\times 10^{10}$ &14.71 &  &  & 0.0241  & 3.81&S0 \\ \vspace{-0.1cm}
& & & & 25.25 & 88 \\
125938.321p275913.89 & PGC 44636 &9.96$\times 10^{9}$ &15.32 &  &  & 0.0384 & 3.50&S0/a \\
\\
130018.873p280033.38 \vspace{-0.1cm}&- & 2.98$\times 10^{9}$ & 16.70 &  &  & 0.0277  &3.29 & S0\\ \vspace{-0.1cm}
& & & & 26.36 & 66 \\
130017.641p275915.27 &- & 9.97$\times 10^{8}$&17.62 &  &  &  0.0191& 3.30&  S0\\
\\
130008.003p280442.81$^1$ & IC 4012 &3.51$\times 10^{10}$ &14.25 & 21.59 & 238\tablenotemark{b} & 0.0447 & 2.59&S0 \\
130012.868p280431.74$^2$ & PGC 44723 &2.11$\times 10^{10}$ &14.66 & 14.32 & 119\tablenotemark{c} & 0.0397 & 3.21&S0 \\
130011.143p280354.91$^3$ &- &2.83$\times 10^{9}$ &16.55 & 20.96 & 119\tablenotemark{d} & 0.0405 & 2.12&S0 
\enddata

\tablenotetext{a}{As defined in \citet{2010ApJS..191..143H} using the prefix {\tt COMAi}}
\tablenotetext{b}{Difference between 1 and 2}
\tablenotetext{c}{Difference between 2 and 3}
\tablenotetext{d}{Difference between 1 and 3}
\end{deluxetable*}

By setting a projected distance limit of $r_s< 30h^{-1}$\,kpc we find 54 of 70 galaxies lying in 50 individual pairs. However, if we add the difference in line-of-sight velocity criteria of $\Delta V \leq 300$\,km\,s$^{-1}$ we find a total of 13 galaxies (listed in Table 1) in five pairs and one triple system, all of them contained within the 19 ACS central pointings. These cuts on $r_s$ and $\Delta V$ are similar to those used in the literature for pair count based merger rate estimations \citep{2000ApJ...536..153P,2004ApJ...617L...9L,2005ApJ...627L..25T,2013MNRAS.429.1051C}.
Stellar mass ratios for the selected pairs range from $\sim$1:1 to $\sim$1:3, that is, if they are actually physically related they could evolve into major mergers.

\subsection{Pair likelihood in Coma}

We investigate statistically the likelihood that these five observed close pairs and a close triple system are interacting and will merge in the future, or if instead they are simply chance alignments due to the high density of cluster galaxies in projected phase-space. For this calculation we adopt spherical symmetry. Considering all known Coma cluster members within $R_{200}$ \citep[1.99 $h^{-1}$ Mpc;][]{2007ApJ...671.1466K}, and having SDSS $ugriz$ photometry, equivalent $g-i$ color and $i$-band magnitude cuts to that used to identify our red sequence population in the ACS F814W and F475W imaging are applied. The position angles of these red sequence Coma galaxies are randomized with respect to the center of the X-ray emission from Coma \citep{2003A&A...400..811N}, and their velocities randomized by repeatedly swapping the redshifts of cluster members. This randomization process should model the expected galaxy density of the virialized population of galaxies in the Coma core, in which all resulting pairs are just chance projections along the line of sight. The expected number of galaxy pairs, with the adopted $r_s$ and $\Delta V$ limits, that would be found within the 19 ACS images based on 10,000 randomized Coma red sequence populations is $7.0{\pm}2.2$, including $1.3{\pm}1.1$ triples (or more complex systems).

The predicted numbers are entirely consistent with the observed number of pairs/triplets, indicating that they all could be simply chance alignments. Nevertheless, this calculation does not rule out some of the observed pairs actually being physical ones. In order to test this, we search for evidence of recent or on-going interactions between galaxies belonging to the observed pairs/triplets.

\subsection{Morphological inspection of galaxies}
\label{visual}

The search for evidence of recent galaxy-galaxy interactions in the 5 pairs and one triplet identified in Section \ref{Pair selection} requires the generation of model subtracted images for the 13 member galaxies. In most cases, it is on these residual images only that tidally induced low surface brightness features can be discerned.

The adopted galaxy models come from \citet{2014MNRAS.441.3083W}. They were obtained using GALFIT \citep{2010AJ....139.2097P}, considering up to three Sersic sub-components. Single Sersic profile models were separated into photometric ellipticals and disks, while multi-component models were classified as S0 and E depending on the index of their main component. The morphological classification and Sersic index for each galaxy is given in Table 1. Features generated by recent galaxy-galaxy interactions generally tend to be highly asymmetric, such as tails, arcs, shells, ripples, bridges, and asymmetric spiral arms. In contrast, most internal features generated through internal instabilities tend to be symmetric with respect to the galaxy center or with respect to some reflection axis. These internal perturbations are easily distinguishable in the GALFIT residuals from the highly asymmetric interaction-driven features. Tidal interactions and minor mergers may also induce ``bar-like'' elongated structures, which tend to display asymmetries (e.g., in length, axial ratio and shapes of dust lanes) not seen in internally-induced bars. Unperturbed galaxies should show smooth gradients towards the outer parts of the galaxy and no noticeable asymmetric structures on the residuals, leaving only minor residuals, such as those arising from the boxy or disky profiles present in some ellipticals or from the presence of bars.

Figure 3 displays F814W images for the 13 galaxies in the selected pairs/triplet, the model subtracted image, and residuals after subtraction of the galaxy after rotation by $180^{\circ}$. Model residuals are smoothed to highlight medium and large-structures. To facilitate the diagnoses from the residual images after galaxy rotation, the foreground and background bright sources are masked. After model subtraction, the presence of structures such as bars and disk
or boxy light profiles can be observed in some of the residual images.
These are all symmetric with respect to the galaxy center, and are thus
likely to be a result of internal processes (e.g. bars). Examples of residuals caused by bars can be seen for IC 4030 and PGC 44636. The boxy profile of IC 4033 is revealed on the corresponding residual image. In all cases, the structures observed in the residuals appear to be symmetrical with respect to the galaxy center, consistent with the expectations for inner galaxy structures.
We have also examined a pair that did not qualify to be included in Table 1, NGC 4898A/B but that is of particular interest, because of the small projected separation of 2.59 h$^{-1}$ kpc and comparable F814W magnitudes (13.42 and 14.38) of the two galaxies. Although the difference in their radial velocities is 532\,km\,s$^{-1}$, higher than the cut employed here, it is lower than the average difference between the pairs selected by projected distance  ($\sim 1100\,{\rm km\,s}^{-1}$). Still, it is large enough to make coalescence unlikely. A visual inspection of the images of this projected pair reveals high overlapping, while the residuals of the GALFIT models reveal asymmetric structures on both galaxies. Nevertheless, these features cannot unmistakenly be regarded as result of an ongoing interaction since the correct modeling becomes more difficult to achieve when the two galaxies overlap. 

In conclusion, for the candidate bound pairs/triplet we do not find indications of low-surface brightness features attributable to recent galaxy-galaxy interactions. 

\subsection{Asymmetry of galaxies in pairs}

The asymmetry parameter $A_{abs}$ is measured for the 13 galaxies in the pairs/triplet following the procedure by \citet{2003ApJS..147....1C} where the intensity of the galaxy and a $180^{\circ}$ rotated image of itself is compared pixel to pixel.
Values for $A_{abs}$ range from 0 for a completely symmetric light distribution to 1 for one that is completely asymmetric. A correction for uncorrelated noise from the background is applied computing the asymmetry parameter for a synthetic area of the same size and rms noise measured close to the galaxy. Sky level subtraction and masking of fore/background sources is applied in order to minimize the effect of non-galactic sources. Typical values for unperturbed early-type galaxies range from 0.01 to 0.1 while irregular and starburst galaxies have been found to have values of 0.2-1.0 \citep{2003ApJS..147....1C, 2012MNRAS.419.2703H}. 

The asymmetry parameters determined for these 13 galaxies are listed in Table 1. Their parameters lie in the 0.02 - 0.06 range, corresponding to unperturbed galaxies, a result consistent with the conclusions of Section \ref{visual}.

\section{Dry merger rate}

The merger timescale for a given number of physical pairs can be estimated using the formula by \citet{2008MNRAS.391.1489K}, which considers typical stellar masses and distances between pair members. The merger timescale is given by 
$$ \langle T_{\mbox{merge}} \rangle = 3.2\mbox{ Gyr}\frac{r_s}{50\mbox{kpc}}\left(\frac{M_*}{4\times 10^{10}M_{\odot}} \right)^{-0.3}\left(1+\frac{z}{20} \right)$$
In section 3.1, we selected 13 galaxies (18.5 \% of the complete sample) complying with the adopted $r_s$ and $\Delta V$ limits, whose median mass and projected separation are $2.11\times 10^{10}M_{\odot}$ and $21.275$ h$^{-1}$kpc, respectively. If, tentatively, it is assumed that these 13 galaxies are actually in interacting systems, then using the above formula a merger timescale of $1.65$ Gyr is obtained, that would lead to a nominal dry merger rate of 11.2\% per Gyr in the Coma core.

However, the visual inspection and asymmetry determination conducted for these 13 galaxies do not provide evidence that they are in interacting systems. This null result, allows nevertheless an estimation of the dry merger rate by using binomial statistics. We follow the procedure by \citet{2003ApJ...586..512B}, where the $\pm$1$\sigma$ range of acceptable values for the pair fractions are defined as a function of the sample size and pair fraction. From our finding of the number of pairs (n=0) for a sample size of 70 galaxies (N=70) we find an upper limit for the merger fraction of $\sim 2.5\%$. Considering the merger time-scale of $1.65$ Gyr estimated above, we obtain a 1-sigma upper limit for the major dry merger rate of $\sim1.5\%$ per Gyr within the Coma cluster core. 
This is not sufficient for dry mergers to account for the red sequence evolution inside clusters \citep{2012ApJ...753...44S},  consistent only with growth rates of $<10\%$ since $z=1.5$ as derived by several previous studies \citep{2007AJ....133.2209D,2008ApJ...686..966M}.

\section{Post-merger fraction}

\begin{figure}
\begin{center}
\includegraphics[width=0.3\textwidth]{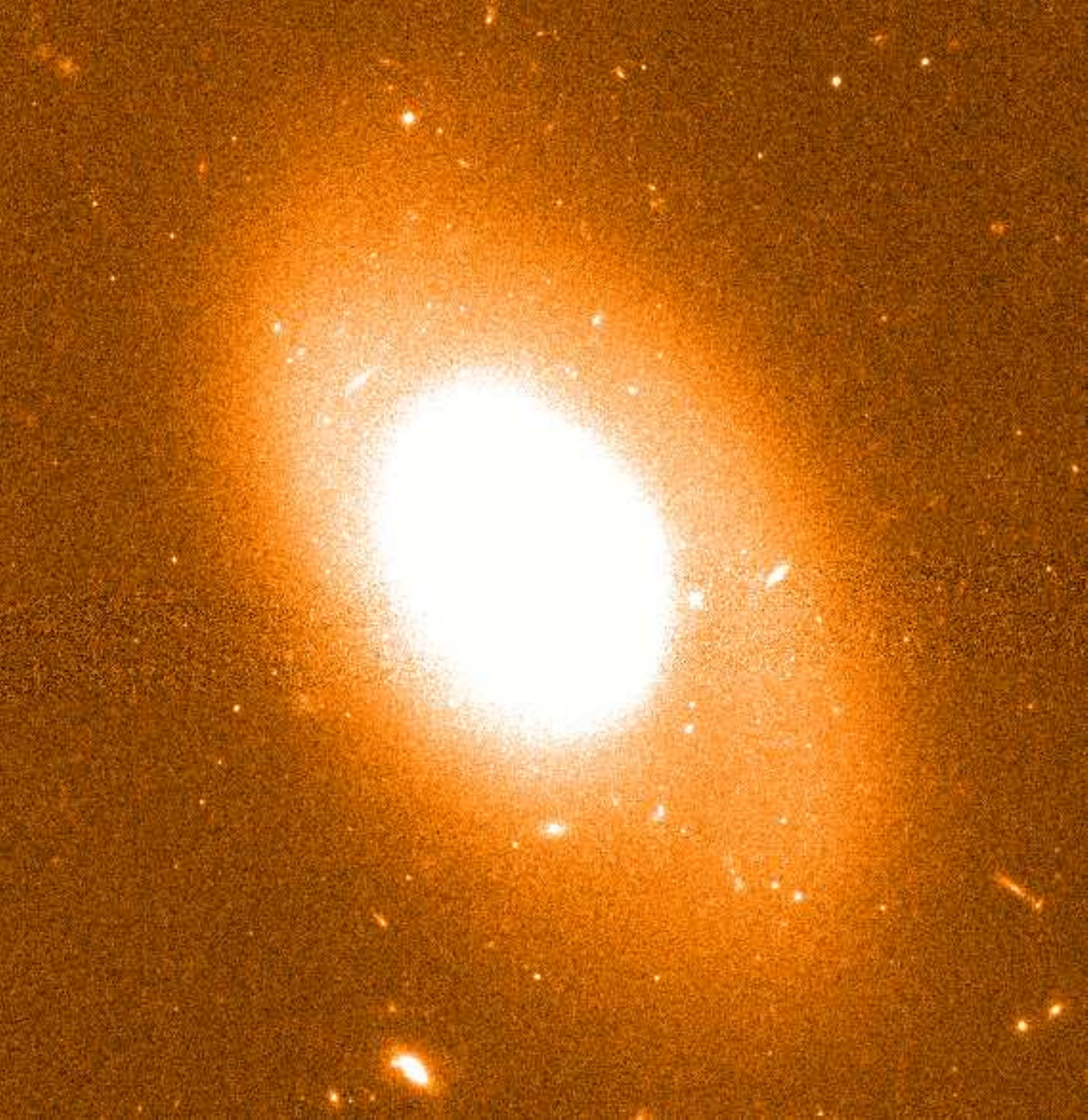}
\includegraphics[width=0.3\textwidth]{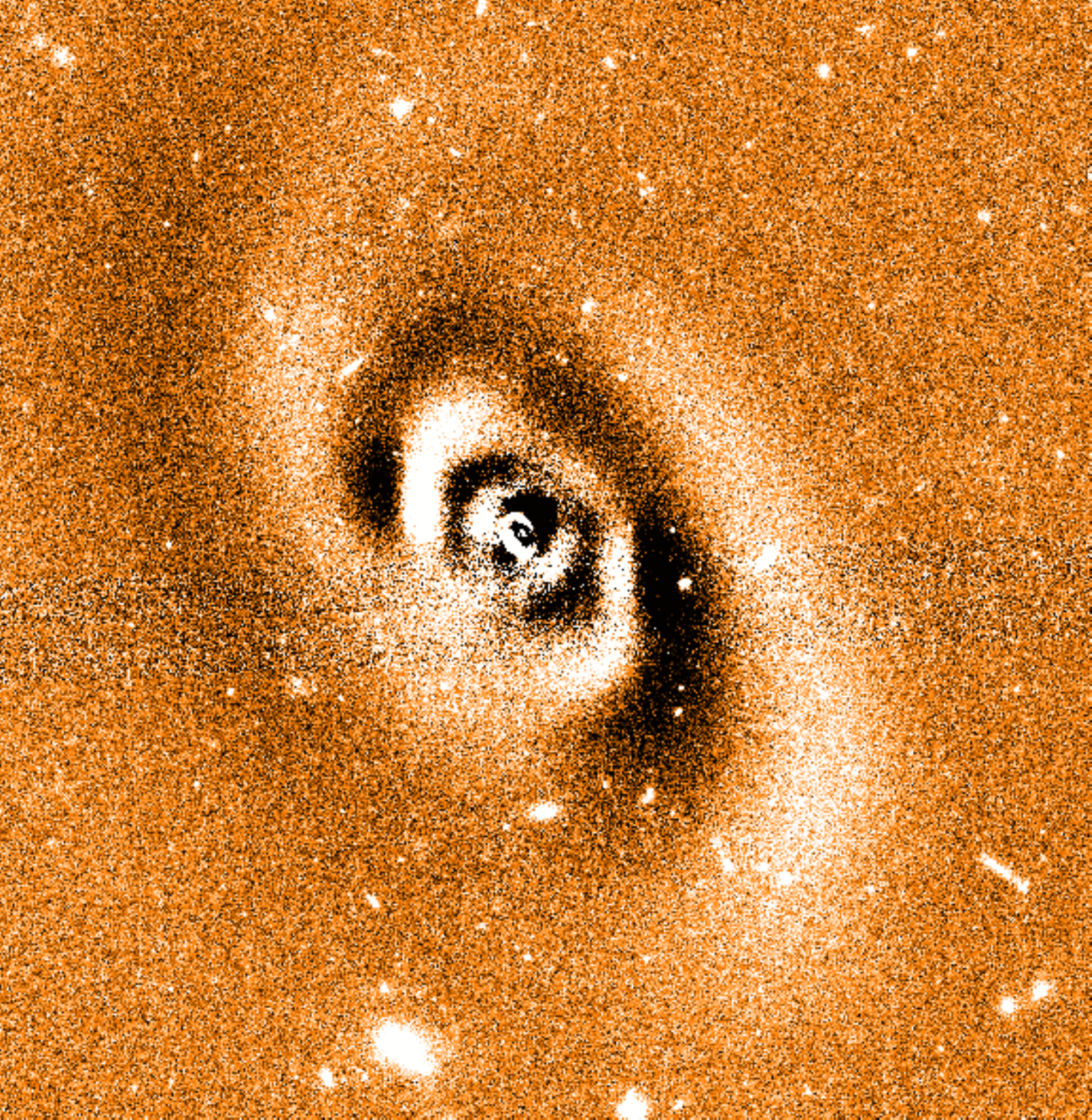}
\caption{On top, F814W image of the only candidate post-merger galaxy IC3973. The lower panel shows the residuals after subtraction of the model. A long, clockwise arm can be observed coming from the south-west.}
\end{center}
\end{figure}

The complete sample of red sequence galaxies is inspected to estimate the fraction which show signatures of the coalescence of two galaxies, such as ripples, tidal structures, halo discontinuities, shells, and other structures unrelated to the presence of a companion. Using images to a SBL of $\sim$26.5\,mag\,arcsec$^{-2}$, only 1 out of the 70 RS galaxies is classified as a post-merger. The only remnant candidate found is IC\,3973, which is previously identified as a member of one of the five pairs and the triple system in Sect. \ref{visual}. This galaxy is not identified as a member of an on-going merger since the observed feature does not appear to be related to the presence of the projected partner, and the partner itself does not show any evidence of perturbation. The asymmetry parameter determined for this galaxy is 0.0419, a value still small for a perturbed galaxy; we believe that this parameter does not actually reflect the merger remnant character detected visually due to the low surface brightness of its external halo compared with the galaxy nucleus. In the direct F814W image of this galaxy (Fig. 4, upper panel), the outer halo of the galaxy appears to be rotated and shifted with respect to the inner bright core, while in the residual image (lower panel), a curved extension is apparent coming out clockwise from the low right corner of the galaxy halo.

Finding only one candidate merger remnant galaxy, implies a post-merger fraction in the Coma cluster core of $\sim 1.4\%$. This result is in agreement with the estimate (3\%) of \citet{2012AJ....144..128A}, to similar surface brightness limits. However, our value is much smaller than the $\sim 25\%$ mean fraction determined by \citet{2012ApJS..202....8S},  even within $\sim 0.2 R_{200}$ (their Fig. 14), in four $z \lesssim 0.1$ Abell clusters using images with a SBL of $\Sigma_r \sim 30$\,mag\,arcsec$^{-2}$. 

\section{Conclusions}
On the one hand, by combining the identification of close pairs with the requirement of galaxy asymmetries, we find no evidence for major on-going mergers in a spectroscopically complete sample of 70 red sequence galaxies within $\sim 0.5 R_{200}$ from the center of Coma and derive an upper limit to the dry merger rate of $\sim 1.5$\% per Gyr at the 1-$\sigma$ sigma level. This rate is not sufficient for dry mergers to account for the red sequence evolution inside clusters.

On the other hand, we find that from the 70 galaxies in our sample only one shows evidence of low surface brightness features identifiable as the remnants of a past merger or interaction, yielding a post-merger fraction of 1.4\% within a projected distance of $\sim0.5 R_{200}$ from the Coma center. Although the Coma brightest member (NGC 4889) is not in our sample, it actually is a red sequence galaxy sitting in the cluster center, and interestingly, it has been found to contain a system of shells identifiable to a minor ($\sim$1/100) merger \citep{2013ApJ...773...34G}. If NGC 4889 would have been part of our sample, presumabley it would have been counted as a galaxy with tidal signatures, implying a larger post-merger fraction, of $\sim$ 2.8 \%. There is however a relevant caveat that derives from the work of \citet{2013ApJ...773...34G} on NGC 4889, i.e. that an observation alone of tidal signatures in a galaxy may sometimes be the consequence of a very minor merger.

The small post-merger fraction we observe is consistent with similar results such as the one by \citet{2012AJ....144..128A}, where $\sim 3\%$ of a large sample of early-type galaxies in clusters ($0.04 < z < 0.15$) have evidence of tidal features found on imaging with surface brightness limits comparable to those of the HST imaging employed by us in this study. However, it is puzzling that the post-merger fraction we observe is a factor of 10 lower than the one measured by \citet{2012ApJS..202....8S} in four $z \lesssim 0.1$ clusters. This discrepancy merits further investigation with due consideration to differences in survey limits and cluster evolutionary stage.

\acknowledgments

LEC and JPC received partial support from the CONICYT Anillo project ACT-1122. LEC thanks support from the Center of Excellence in Astrophysics and Associated Technologies (PFB06). JPC acknowledges CONICYT/PCHA/MagisterNacional/2014 - folio 22141888. CPH was funded by CONICYT Anillo project ACT-1122. SJ acknowledges support from NSF grant AST-1413652 and the NASA/JPL SURP program. This research made use of the NASA/IPAC Extragalactic Database (NED) which is operated by the Jet Propulsion Laboratory, California Institute of Technology, under contract with the National Aeronautics and Space Administration. We thank the referee for suggestions that contributed to the improvement of the paper.

\end{document}